\documentstyle[11pt,adassconf]{article}  

\begin{document}   

%
%
%

\paperID{P4.14}

%
%
\title{ Predictive Mining of Time Series Data in Astronomy}

%
%
%

\author{ Eric Perlman, Akshay Java}
\affil{Joint Center for Astrophysics, University of Maryland, Baltimore County,
1000 Hilltop Circle, Baltimore, MD  21250 }

%
%

\contact{ Eric Perlman}
\email{ perlman@jca.umbc.edu}

%
%
%

\paindex{ Perlman, E.}
\aindex{ Java, A.}     

%
%



\begin{abstract}          

We discuss the development of a
Java toolbox for astronomical time series data. Rather than using methods
conventional in astronomy (e.g., power spectrum and cross-correlation analysis)
we employ rule discovery techniques commonly used in analyzing stock-market
data. By clustering patterns found within the data, rule discovery allows one
to build predictive models, allowing one to forecast when a given event might
occur or whether the occurrence of one event will trigger a second. We have
tested the toolbox and accompanying display tool on datasets (representing
several classes of objects) from the RXTE All Sky Monitor. We use these
datasets to illustrate the methods and functionality of the toolbox. We also
discuss issues that can come up in data analysis as well as the possible
future development of the package.

\end{abstract}

%
%

\section{ Introduction}
 
Many types of variable objects exist in the universe, including stars with
predictable behavior (e.g., Cepheids), objects with behavior that is inherently
unpredictable (e.g, AGN), and objects with both predictable and irregular
variability patterns (e.g., X-ray binaries).  Constant monitoring of variable
objects has been a continuing interest in astronomy, beginning with 16th
century astronomer David Fabricius, and extending through history to Herschel,
Leavitt and others.  Today, monitoring is done by a wide variety of techniques,
observers and instruments, from dedicated amateurs, to professional astronomers
interested in intensive monitoring of individual objects, to all-sky monitors
such as the RXTE ASM and BATSE aboard CGRO.  

While they are a new tool, already all-sky monitors have have made important,
if not decisive contributions to solving some of astronomy's most persistent 
mysteries, such as the cosmological origin of gamma-ray bursts and linking
emission regions in AGN.  With major initiatives such as the Large-Area
Synoptic Survey Telescope (LSST) and Supernova Acceleration Probe (SNAP),
all-sky monitors are poised to become a major discovery tool in astronomy. To
maximize the utility of large monitoring programs, it is important to devise
ways of handling large amounts of data in real time and find not only
variability but also predictive patterns among these large data streams. It was
with these goals in mind that we undertook this project.

\section{Clustering and Rule Discovery}

The need to understand time series data is not unique to astronomy:  time
series data exists in a wide variety of fields including geology and
atmospheric science as well as many business applications.  The common problem
is how to efficiently find patterns in the data.  Fourier transform and power
spectrum methods, commonly used in astronomy research, are well suited for
finding patterns with well-determined periodicities (e.g., see Scargle 1997). 
However they may be less helpful for objects with irregular behavior, and are
not optimized to serve as predictive tools.  Our approach  was to employ
clustering and rule discovery, which are optimized for finding  patterns that
do not rely on a regular periodicity.  

\begin{figure}

\plotone{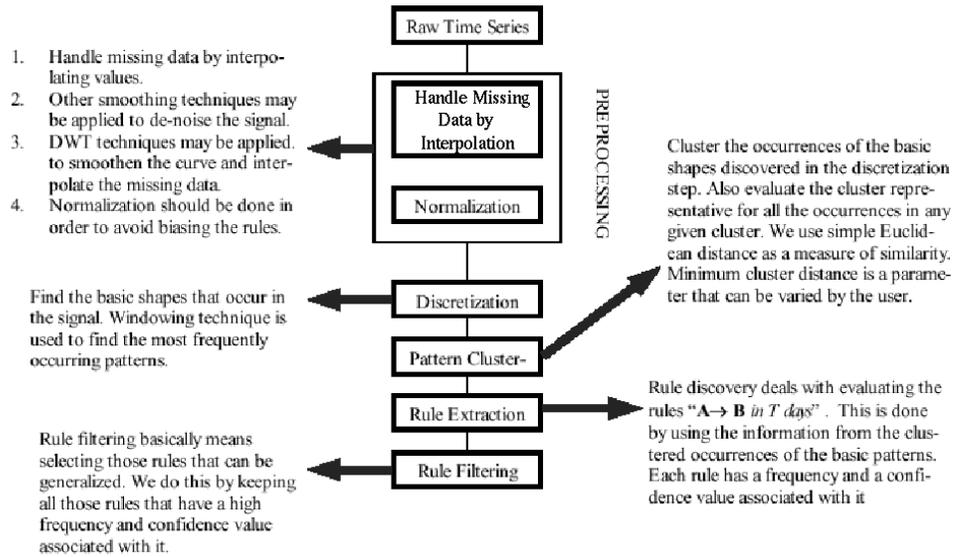}

\caption{A flow chart describing the basic rule discovery algorithm}

\end {figure}

We first attempt to represent the dataset as a collection of patterns, by
sliding a window of size through the time series to get
subsequences.   We then cluster these points using the greedy clustering
algorithm explained in Das, Gunopulos \& Mannila 1997).  Once a good fit is
achieved, these clusters can then be considered as the basic shapes of the time
series, with the entire series composed of superpositions and/or combinations
of these basic shapes. The next and main step of the process  is  to find
interesting probabilistic rules between the clusters in the two time series.
These rules are of the form:  "If a cluster A occurs in time series 1 then we
can say with confidence c  that B will occur in time series 2 within time T". 
Figure 1 displays a flow chart of the basic algorithm used in predictive time
series analysis.

\section {Data Analysis Structure and Usage}

Two goals are possible with this approach, each requiring a slightly different
analysis tree:  (1) to schedule a telescope more intelligently to objects when
they are at their brightest or most interesting stage, or (2) as an analysis
tool to find patterns within a (possibly multivariate) time series (see 
Figure 2).

\begin{figure}

\plotone{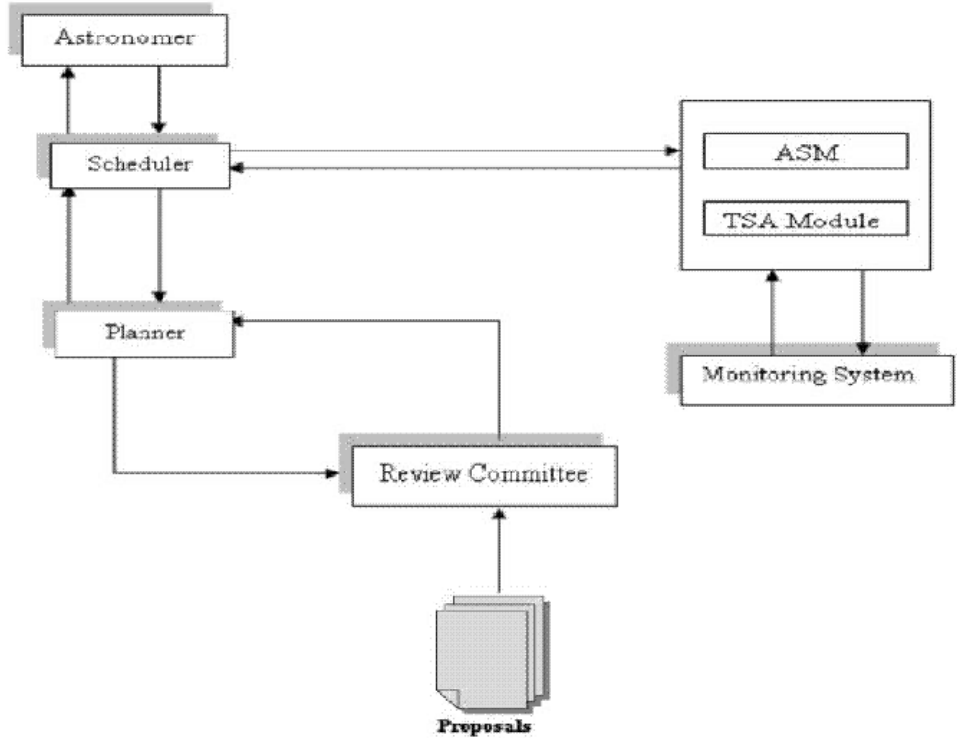}

\caption{Flow charts for the possible use of this algorithm in scheduling a
telescope.}

\end{figure}

Within the time-series analysis (TSA) module, the user can control three main
parameters:  the window size, the minimum cluster distance and the time period
for prediction.  The TSA module analyzes the time series data and performs
trend prediction and rule discovery on both real time and historical data. This
output can either be used directly for data analysis in a paper or to
forecastboth the activity level of an object and the ability of the telescope
or satellite to perform the desired observations.  All of these factors could
be considered in the light of the constraints imposed by individual
investigators  as well as the review committee. The goal of such a process
would be to maximize the scientific  usefulness of all the observations.  If,
for example, a given observation required that a source be above a certain flux
level, such a program might have a better chance of ensuring that this was so
during the observations.

\section{Results, Limitations and Future Work}

To test and validate the toolbox, we used RXTE ASM datasets for various
objects, spanning several different object classes. In Figure 3, we show one
particularly interesting dataset, from SMC X-1.  SMC X-1 shows interesting
periodic behavior on several different timescales:  a long time-period behavior
as well as short time-scale variations superposed upon the long-period
behavior, but occurring only when the object is bright (see e.g., Kahabka \& Li
1999).  The toolkit was found to be particularly effective for predicting
short-period rules.  Longer-period rules can be found by varying the window
size, although possibly at the expense of losing  short time-scale rules as the
window size becomes larger than the rules in question.  We can also find
longer-period rules with smoothing techniques.  

The current implementation of the toolkit consist of a rule inference engine,
programmed in C and a visualizer, programmed in Java.  The current
toolkit has several limitations.  First, the current toolkit can analyze only a
single time series at any time.  This obviously gives it limited functionality,
as many objects are studied with multiple instruments.  Second, it has limited
ability to deal with missing data.  In addition, one could envision multiple
processing options, including the incorporation of dynamic time warping
techniques, which could be particularly useful to spot quasi-periodic 
oscillations.

\begin{figure}
\plotone{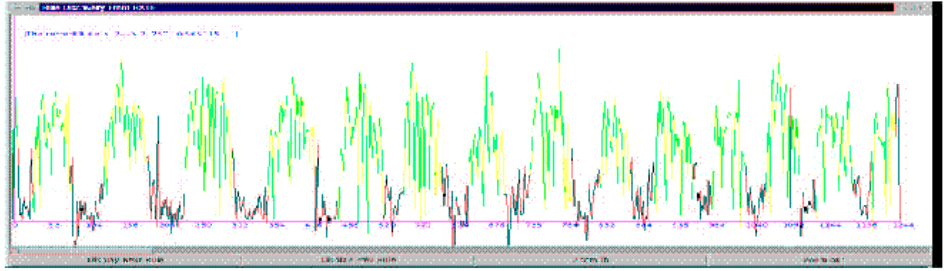}
\caption{ A screenshot of the  Java GUI application that is used to display
the rules generated. Each color represents an occurrence of a basic pattern. As
each rule is cicked on, the GUI shows the rule's frequency 
and associated confidence values.}
\end{figure}

\end{document}